# Ultralow noise microwaves with free-running frequency combs and electrical feedforward

Takuma Nakamura[1,2,*], William Groman[3,4], Qing-Xin Ji[5], Oguzhan Kara[6], Benjamin Rudin[6], Anatoliy Savchenkov[7], Vladimir Iltchenko[7], Wei Zhang[7], Andrey Matsko[7], John E. Bowers[8], Florian Emaury[6], Kerry J. Vahala[5], Scott A. Diddams[3,4], and Franklyn Quinlan[1,4,■]

[1]Time and Frequency Division, National Institute of Standards and Technology, 325 Broadway, Boulder, CO 80305, USA
[2]Department of Electrical Engineering, University of Colorado Denver, 1200 Larimer Street, Denver, CO 80204, USA
[3]Department of Physics, University of Colorado Boulder, 440 UCB, Boulder, CO, 80309, USA
[4]Electrical, Computer and Energy Engineering, University of Colorado, Boulder, Colorado 80309, USA
[5]T. J. Watson Laboratory of Applied Physics, California Institute of Technology, Pasadena, CA, USA
[6]Menhir Photonics AG, Zürichstrasse 130, 8600 Dübendorf, Switzerland
[7]Jet Propulsion Laboratory, California Institute of Technology, Pasadena, CA, USA
[8]Department of Electrical and Computer Engineering, University of California, Santa Barbara, Santa Barbara, CA, USA
*takuma.nakamura@nist.gov; ■fquinlan@nist.gov

**Abstract**

Optically generated microwave signals exhibit some of the lowest phase noise and timing jitter of any microwave-generating technology to date. The success of octave-spanning optical frequency combs in down-converting ultrastable optical frequency references has motivated the development of compact, robust and highly manufacturable optical systems that maintain the ultralow microwave phase noise of their tabletop counterparts. Two-point optical frequency division using chip-scale components and ~1 THz-spanning microcombs has been quite successful, but with stringent requirements on the comb source's free-running noise and feedback control dynamics. Here we introduce a major simplification of this architecture that replaces feedback control of the frequency comb in favor of electronic feedforward noise cancelation that significantly relaxes the comb requirements. Demonstrated with both a high repetition rate solid-state mode-locked laser and a microcomb, feedforward on a 10 GHz carrier results in more robust operation with phase noise as low as -153 dBc/Hz at offsets ≥10 kHz, femtosecond timing jitter, and elimination of the large "servo bump" noise increase at high offset frequency. The system's compatibility with a variety of highly manufacturable mode-locked laser designs and its resilience and straightforward implementation represents an important step forward towards a fully chip-scale implementation of optically generated microwaves, with applications in radar, sensing, and position, navigation and timing.

## Introduction

The spectral purity of an electrical signal is one of the most fundamental and critical parameters for many scientific and industrial measurements, such as sensing with microwave interferometry[1], Doppler radar[2], atomic clocks[3], and precision position, navigation and timing applications[4]. While the spectral purity is often characterized in terms of phase noise, the integrated timing jitter is usually just as critical to system performance[5]. By exploiting the extremely low loss and high quality factors available in the optical domain, optically generated microwave signals have exhibited extremely low phase noise and timing jitter, towards meeting the needs of these applications. Optical frequency division (OFD) in particular, where a low noise optical frequency reference is divided to the microwave domain by way of an octave-spanning optical frequency comb[6–8], has produced extremely low phase noise on a 10 GHz carrier both at high and low frequency offsets. On the other hand, cryogenic and room temperature microwave sapphire loaded cavity oscillators can produce low phase noise only at certain offset frequencies[9,10].

Importantly, the small size, weight, and power (SWaP) requirements of many out-of-the-lab applications of low noise microwaves often preclude the use of an octave-spanning optical frequency comb. Widespread adoption of optically derived microwaves will therefore require the development of compact, robust and highly manufacturable optical systems that can deliver phase noise better than traditional microwave sources of a similar footprint. Whereas sub-10 cm-scale mode-locked lasers can generate extremely low free-running noise[11], compact and chip-scale sources largely rely upon on 2-point OFD (2p-OFD) that requires a comb bandwidth of only ~1 THz. This can be produced with a wide variety of comb sources, such as a semiconductor or solid-state mode-locked laser, an electro-optic modulation frequency comb or a microcomb. In 2p-OFD, two THz-spaced optical modes of the comb are anchored to a single optical reference[12]. Feedback control to the comb then divides the frequency separation $Nf_{rep}$ of these modes to generate a low noise microwave frequency carrier at the fundamental comb spacing $f_{rep}$. Using state-of-the-art microcombs, on-chip lasers and compact optical cavities, several groups have recently demonstrated compact 2p-OFD architectures, generating microwave signals whose phase noise reaches approximately -140 dBc/Hz at 10 kHz offset when scaled to a 10 GHz carrier[13–16]. Challenges remain in the creation of robust and manufacturable systems, however, including the availability of suitably low noise comb sources and the complexity of stabilization architectures that provide durable and long-lasting feedback control.

Here we forego comb feedback in favor of electronic feedforward noise cancellation, demonstrating a major simplification of the 2p-OFD architecture that simultaneously improves performance and robustness. We demonstrate this technique using two compact comb sources. With a commercial 10 GHz repetition rate solid-state mode-locked laser, we achieve phase noise on a 10 GHz carrier that is below -115 dBc/Hz at 100 Hz offset and remains below -150 dBc/Hz for offset frequencies larger than ~6 kHz. Importantly, this low and flat phase noise level would be unachievable with standard feedback control, even assuming a comb with an ultrafast actuator with bandwidth exceeding 1 MHz[17–19]. With a microcomb module operating at 20 GHz, the microwave phase noise scaled to a 10 GHz carrier reaches -120 dBc/Hz at 100 Hz offset and remains below -145 dBc/Hz for offset frequencies larger than 10 kHz. This result is comparable to the best microwave phase noise results to date with a microcomb. Moreover, for both comb sources, there is no "servo bump" noise increase associated with feedback control that often peaks 20 dB or more above the phase noise floor at frequencies near the edge of the feedback bandwidth. As a result, the integrated timing jitter (1 MHz to 10 Hz) for the mode-locked laser microwave is reduced to only 2.9 fs. For the microcomb, the integrated jitter is 2.5 fs, lower than the jitter of any other microcomb-generated microwave or millimeter wave signal of which we are aware. Additionally, since there is no feedback loop, the comb cannot unlock or cycle-slip, extending the duration of low noise operation – with the microcomb we operate for days on end, and with the solid-state mode-locked laser low noise operation extends indefinitely. This demonstration shows that the feedforward architecture can expand the diversity of compact comb sources available to low noise microwave generation, reduce the timing jitter of 2p-OFD, and simplify the system architecture for manufacturable systems requiring long-term operation.

**Results**

*System concept*

Our feedforward scheme is shown conceptually in Fig. 1. Just as with the standard 2p-OFD, two independent continuous-wave (CW) lasers, typically with frequency separation in the range of 0.5 to 2 THz, are stabilized to the same optical reference. While not strictly necessary, stabilizing both lasers to the same reference is preferred, since the microwave output will reject noise from the reference that is common to both lasers. To divide the CW laser frequency difference to a microwave carrier, these lasers are first heterodyned with nearby modes of the frequency comb. The two heterodyne beats are then combined in a frequency mixer, generating an electrical signal with noise that is given by $N \times \delta f_{rep} + \delta(\nu_{CW2} - \nu_{CW1})$, where $\delta f_{rep}$ represents fluctuations in the comb repetition rate, $\delta(\nu_{CW2} - \nu_{CW1})$ is the relative fluctuation of the two CW lasers at frequencies $\nu_{CW1}$ and $\nu_{CW2}$, and $N$ is an integer. Note that this step not only removes common-mode noise from the optical frequency reference, it also removes any noise from the comb's offset frequency. In the standard 2p-OFD feedback configuration, the mixer output is used to actuate on the comb using a phase lock loop. Ideally, photodetection of the phase-locked frequency comb output will result in a microwave signal with noise given by $\delta(\nu_{CW2} - \nu_{CW1})/N$, that is, the relative noise of the two CW lasers divided by the comb span. However, the feedback loop's finite gain and phase delays can result in unsuppressed comb noise. This is most prominent near the edge of the servo loop bandwidth with the appearance of a servo bump. Moreover, cycle-slips, where phase coherence is momentarily lost, or a more complete loss of phase lock requiring user intervention, can limit long-term operation.

In our feedforward method, the mixer output is instead electronically frequency-divided by $N$ to generate a signal with frequency noise given by $\delta f_{rep} + \delta(\nu_{CW2} - \nu_{CW1})/N$. Separately, the free-running comb output is photodetected to generate a microwave signal with noise $\delta f_{rep}$. The microwave signal is then frequency-mixed with the output of the frequency divider, thereby subtracting the comb repetition rate noise from the microwave carrier, leaving only $\delta(\nu_{CW2} - \nu_{CW1})/N$. The advantage is that no feedback control dynamics come into play, leading to large, broadband suppression of $\delta f_{rep}$ with robust, long-term operation and without a servo bump.

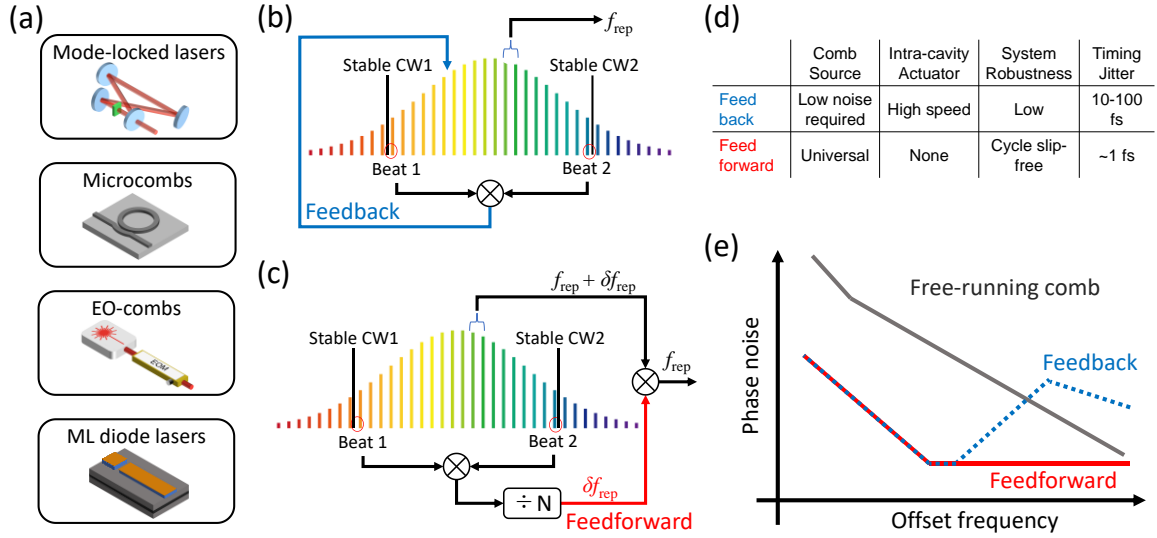

Figure 1. **Schematic concept of low noise microwave generation using a free-running comb source and 2-point OFD.** (a) Potential comb sources for proposed method. EO-combs; Electro-optic combs, ML diode lasers; mode-locked semiconductor diode lasers. Virtually any comb source can be used because low free-running noise and high-speed modulators are not necessary. (b) Conventional 2p-OFD. (c) Proposed 2p-OFD with free-running comb and feedforward. (d) Table comparing conventional 2p-OFD against the proposed method. (e) Projected phase noise plots with conventional 2p-OFD and the proposed method.

The mixer output signal that corresponds to the subtraction of the noise of the two input signals is shifted from the original comb repetition frequency, but, importantly, still oscillates at a microwave frequency. For example, in the experiments described below, the frequency shift is several 10s of MHz on a 10 GHz or 20 GHz carrier. We find this level of frequency shift to be quite convenient, since it is large enough to be readily separated with a microwave filter while remaining comfortably within the same designated microwave band.

*Experimental Setup and Phase Noise Results*

The experimental setup for 2p-OFD with feedforward correction is shown in Fig. 2. We conducted separate experiments with two comb sources – a commercial mode-locked laser operating directly at 10 GHz repetition rate and output power ~30 mW from Menhir Photonics, and a micro-resonator comb operating at 20 GHz with output power ~100 μW[14]. For both comb sources, the optical spectrum spanned the ~1.16 THz gap between two CW lasers operating at 1550 nm and 1560 nm. The comb and CW laser spectra are shown in the upper right of Fig. 2. For each comb source, the comb output was amplified and split into three branches. Two branches were spectrally filtered and heterodyned against the CW lasers. The CW lasers are stabilized to a common compact ultralow expansion (ULE) cavity by the Pound–Drever–Hall (PDH) locking method[13,20]. Special care was taken to reduce the residual noise in the PDH locks to support ultralow microwave phase noise. In particular, the PDH feedback bandwidths are significantly enhanced by using electro-optic modulator (EOM) feedback[21], such that, after electrical frequency division, the in-loop noise supports microwave phase noise below -160 dBc/Hz in the offset frequency range of 10 kHz to 1 MHz. See Methods section for more information on the CW laser stabilization. A commercial digital frequency divider (divide by 58) is used to divide the mixed heterodyne signal output. Only for the mode-locked laser setup, a purpose-built regenerative divider[22] (divide by 2) was added after the commercial divider to overcome the divider's additive phase noise floor of -153 dBc/Hz, improving the achievable phase noise floor to approximately -159 dBc/Hz. For each comb source, the final division ratio matched the number of comb lines spanned by the CW lasers, $N = 116$ for the mode-locked laser and $N = 58$ for the microcomb.

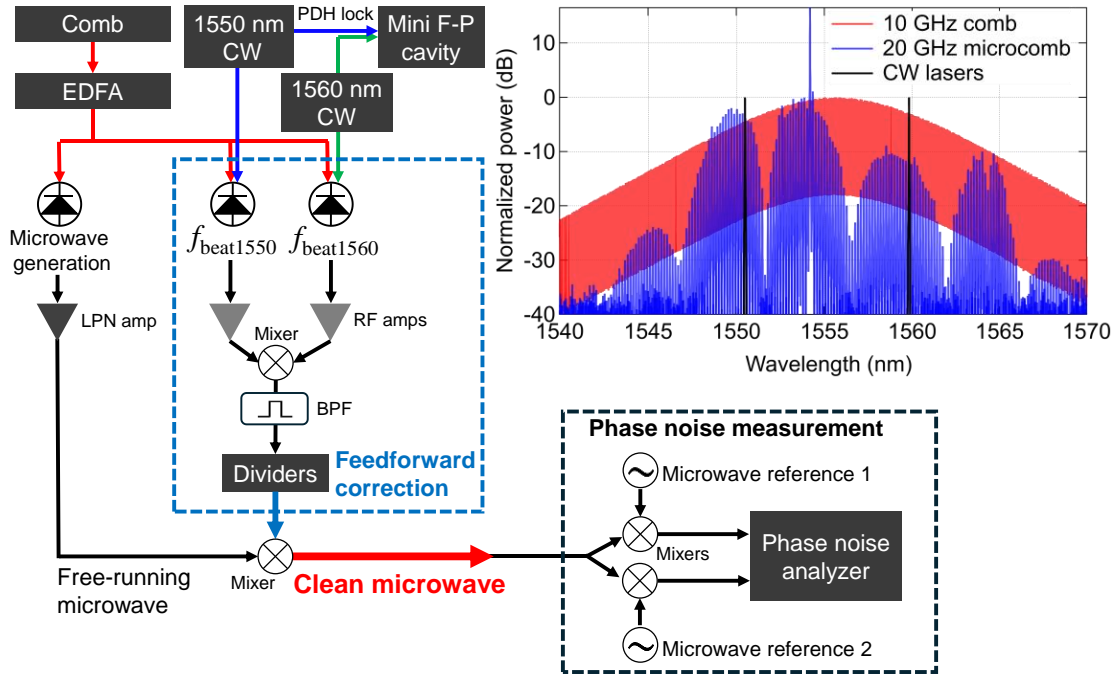

Figure 2. **Detailed experimental setup for low noise microwave generation using 2-point OFD and feedforward correction.** EDFA; erbium-doped fiber amplifier. LPN amp; low-phase noise amplifier, RF amps; radio frequency amplifier, BPF; bandpass filer.

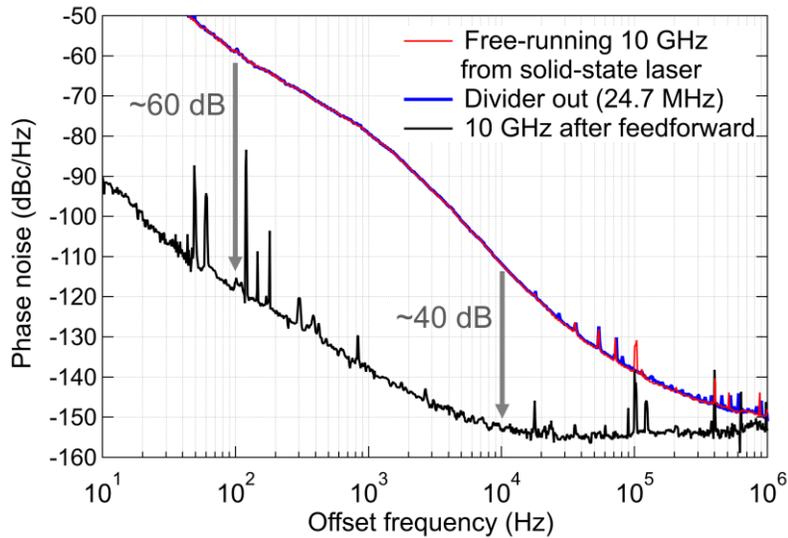

Figure 3. **Microwave phase noise using the solid-state laser.** The red trace shows the free-running phase noise of the photodetected pulse train at 10 GHz. Feedforward correction using the divider-out signal (shown in blue) results in the low noise ~10 GHz output shown in the black trace.

The remaining branch from the optical amplifier illuminates a high-speed photodiode for microwave generation. The microwave signal is amplified by a low-phase noise amplifier and then mixed with the divider output, resulting in the feedforward-corrected microwave carrier. This low noise microwave signal is then directed to a measurement system for cross-correlation phase noise analysis. Here, the low noise microwave is split into 2 branches, and downconverted to ~20 MHz by two independent microwave oscillators. As these reference oscillators exhibit higher phase noise than the microwave signal under test, the downconverted signals are then cross-correlated using a commercial phase noise analyzer. Cross-correlation rejects the noise of the reference oscillators and downconverting mixers, such that the low phase noise microwave output can be measured[23].

The 10 GHz phase noise results from the mode-locked laser are shown in Fig. 3. The red line shows the free-running 10 GHz phase noise from the directly photodetected frequency comb output, and the blue line shows the divider output at 24.7 MHz that is used for feedforward correction. The free-running 10 GHz phase noise exactly matches the 24.7 MHz signal, as required for noise cancellation after mixing. The black trace is the feedforward-corrected 10 GHz signal. At 100 Hz offset, feedforward correction reduces the phase noise by ~60 dB, down to the limit imposed by the non-common noise of the cavity-stabilized CW lasers. The phase noise floor reaches -155 dBc/Hz at 30 kHz offset and remains below -150 dBc/Hz out to the 1 MHz limit of our measurement range. Importantly, no servo bump is observed. As this phase noise was obtained from a completely free-running mode-locked laser, there are no cycle-slips or loss of lock often present in standard feedback control. The system works continuously as long as the individual heterodyne beat frequencies between the comb modes and the CW lasers remain within the system bandpass frequency range. This condition is easily met: with the mode-locked laser used here, the beat frequencies drifted less than 1 GHz in a week, compared to the bandpass range that approaches 5 GHz (one-half the repetition rate). This robustness allowed us to perform a continuous phase noise measurement of more than 90 hours duration with no user intervention.

As with the standard 2p-OFD using feedback control, an important parameter of feedforward noise correction is the signal-noise-to-ratio (SNR) of the individual beatnotes between comb modes and the CW lasers. For a 1 THz span and 10 GHz microwave generation, frequency division reduces the noise on the beatnotes by 40 dB ($20\times\log(1\ \text{THz}/10\ \text{GHz})$). Therefore, an SNR >110 dB/Hz is necessary to support -150 dBc/Hz on the microwave output. Assuming shot noise-limited detection with a strong CW laser as a local oscillator, >100 nW/mode should be sufficient to achieve the required SNR, corresponding to a total comb power of only several tens of μW spread across ~100 comb lines. We explored the SNR limitation by intentionally introducing loss between the EDFA and one of the photodiodes used for beatnote detection. Note that the optical power of the comb for the microwave generation and the beat detection at 1560 nm remained the same throughout the measurement. Results of back-to-back 10 GHz phase noise measurements are shown in Fig. 4. The inset of Fig.

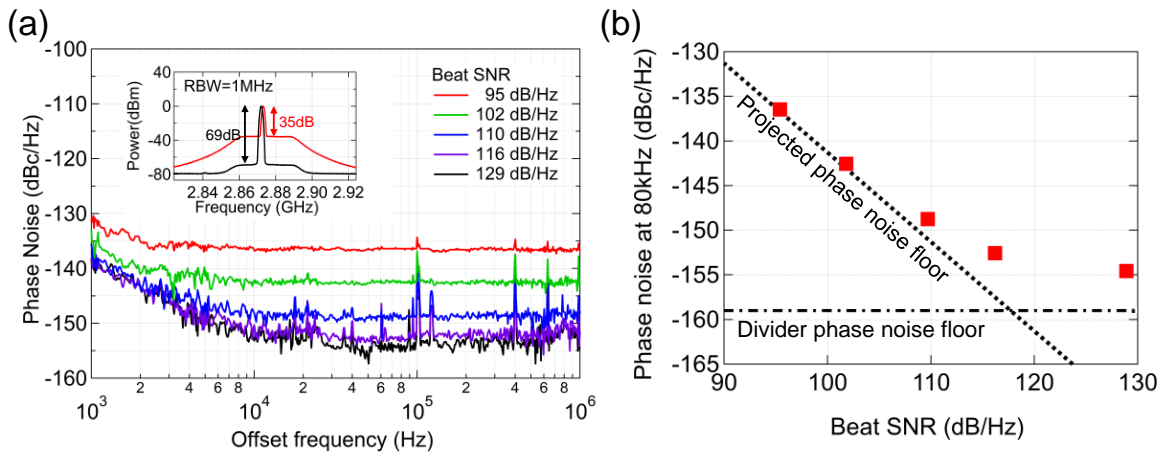

Figure 4. **Determination of the SNR-limited noise floor.** (a) 10 GHz phase noise measurements with different beat SNR. (Inset) Representative mixed heterodyne signals when the SNR was 95 dBc/Hz and 129 dBc/Hz (b) Summary of the SNR effect, using the phase noise at 80 kHz offset for the comparison. To suppress the measurement noise uncertainty, all points represent the averaged noise around 80 kHz offset.

4(a) shows the signal input to the frequency divider, whose SNR is given by that of the combined individual comb-CW laser heterodyne beats. The dependence of the phase noise on the beatnote SNR is shown in Fig. 4(a), and Fig. 4(b) compares the measured phase noise to the expected values. With SNR reduced below 110 dBc/Hz, the measured phase noise follows the projected phase noise floor very well. Higher SNR values approach the phase noise floor of the divider.

To demonstrate the versatility of our method and potential toward a fully integrated system, we also generate stable 20 GHz microwaves using a microcomb source[14]. The 100 µW output power from the microcomb includes residual pump light as shown in the blue spectrum in Fig. 2. We tune the microcomb to a double-soliton state, resulting in a modulated optical spectrum. This provides the highest comb output power, highest 20 GHz power and lowest shot noise floor, leading to the lowest attainable phase noise floor. Note, however, that we could generate a low noise microwave output with any coherent comb state of the microcomb, although the phase noise floor level will vary. As we only replaced the comb source, the feedforward scheme is nearly the same as above, including the CW lasers and reference cavity locking, with the only exception being the heterodyne beat frequencies and the frequencies of other signals derived therefrom. Phase noise results are shown in Fig. 5. For ease of comparison to the mode-locked laser results and other results in the literature, the phase noise level is scaled to a 10 GHz carrier in post processing. On a 10 GHz carrier, the phase noise floor at 10 kHz offset is -145 dBc/Hz and is flat at the level of -146 dBc/Hz from 30 kHz to 1 MHz. At 100 Hz offset, the free-running microcomb noise is rejected by 80 dB. Remarkably, >45 dB suppression of the free-running microcomb noise was achieved at 10 kHz offset. Compared to our previous 2p-OFD microcomb result[13], there is no feedback servo bump, resulting in a >20 dB noise improvement near 100 kHz offset. The phase noise floor of the microcomb-based system is higher than that of the solid-state mode-locked laser, which we attribute to the lower comb power directly from the oscillator that impacts the noise floor of the photodetected 20 GHz signal. Importantly, the SNR of the heterodyne beatnotes is measured to be 113 dB/Hz, a level high enough to support <-150 dBc/Hz floor at 10 GHz. With higher power from the microcomb[24], we anticipate the phase noise reaching -155 dBc/Hz, comparable to our results with the mode-locked laser. We continuously maintained low noise output for several hours, limited only by the longevity of the comb state. Moreover, the beat frequencies with the CW lasers were highly reproducible even after restarting the system after a one-week interruption.

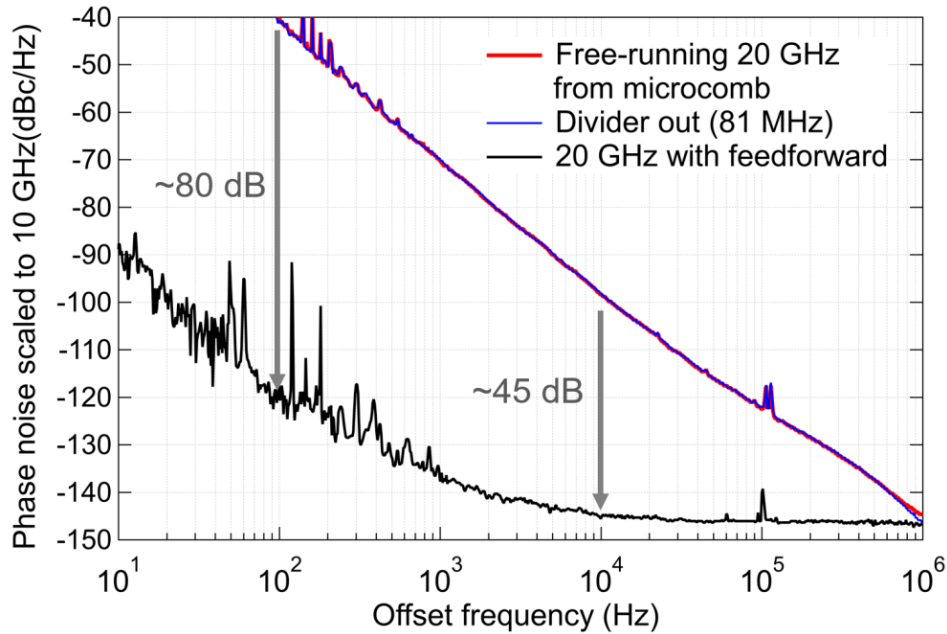

Figure 5. **Microwave phase noise using the micro-comb.** While the micro-comb operates at 20 GHz repetition rate, all phase noise shown here is scaled to 10 GHz for ease of comparison. The free-running micro-comb repetition rate noise is shown in the red trace, the divider output at 81 MHz used for the feedforward correction is shown in blue, and the feedforward-corrected microwave output is shown in black.

**Discussion and Conclusion**

The low microwave phase noise of our feedforward-corrected mode-locked laser and microcomb systems is comparable to state-of-the-art systems of similar footprint, but with two important distinctions: the simplicity and robustness of the feedforward architecture and its lower integrated timing jitter. Feedforward correction provides large rejection of the frequency comb noise without any feedback. This simplifies the system, provides for a larger variety of comb sources capable of low noise microwave generation, eases frequency comb manufacturing requirements, and, as discussed above, allows for continuous operation for indefinite periods of time. The relative delay between the directly photodetected and the feedforward correction signal can limit the level of noise suppression (see Methods). However, even relative delays as large as 100 ns will provide phase noise suppression of 45 dB at 10 kHz offset. We measure the delay of the frequency divider as only 1-2 ns, such that long compensatory delays are not required. Also, as discussed in Methods, the divider is capable of coherently dividing input signals with phase noise 40 dB higher than that of the mode-locked and microcomb sources used here, providing a path for the use of much noisier comb sources, such as mode-locked semiconductor diode lasers. We also note that, while our feedforward noise cancellation method shares similarities with transfer oscillator techniques[25,26], our method doesn't require an octave-spanning comb and a complicated digital synthesis chain that unavoidably adds >100 ns delay. Still, we achieve a comparable or lower microwave phase noise floor.

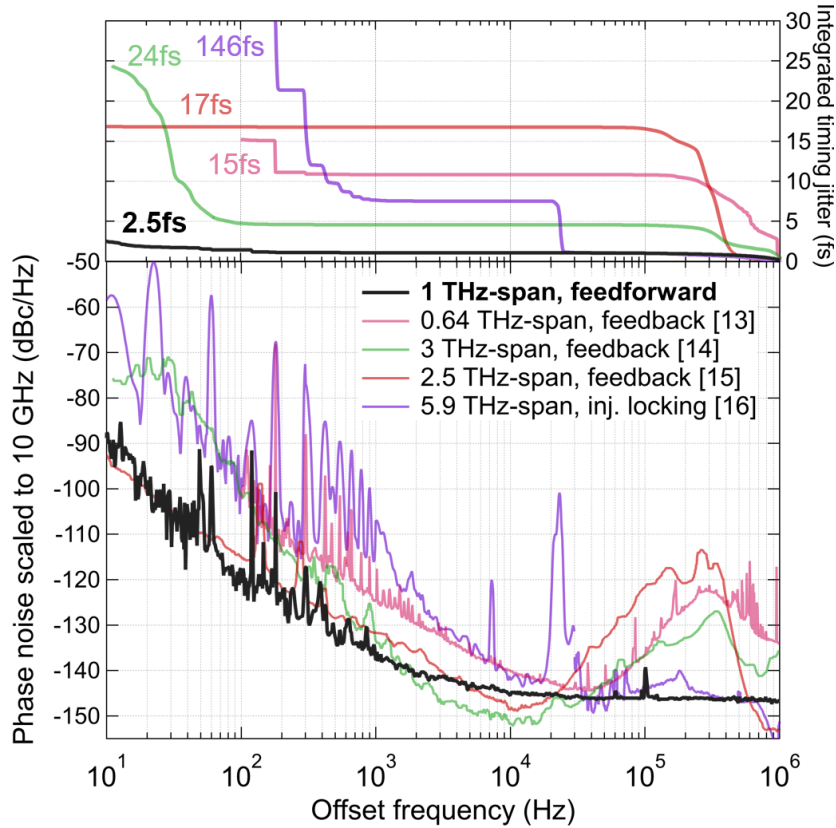

Figure 6. **Phase noise and timing jitter comparison of systems using chip-scale frequency comb sources.** Bottom: Phase noise spectra from several recent publications on low noise microwave and millimeter-wave generation studies using microcombs[13-16]. Phase noise traces are taken from public repositories that accompanied the respective paper publications. The optical frequency span of each microcomb is noted in the legend. Top: integrated timing jitter for the same set of phase noise results. For the 0.64 THz-span comb, the integration range is 1 MHz to 100 Hz. All other timing jitter results are for an integration range of 1 MHz to 10 Hz.

While phase noise at specific offset frequencies is often used to characterize the performance of a microwave source, the integrated timing jitter can be just as critical, such as for high-fidelity digital sampling[5]. In Fig. 6 we compare the phase noise and integrated timing jitter of several recent reports on low noise microwave generation with ~THz-span, chip-scale optical frequency comb sources. At 2.5 fs, the integrated timing jitter from 1 MHz to 10 Hz of our microcomb-based microwave signal with feedforward correction is well below that of comparable systems using feedback control. This jitter reduction is largely due to the lack of a servo bump, though a clean spectrum free of spurious tones and low close-to-carrier noise reduce the integrated jitter as well. Using feedforward, the integrated jitter of the mode-locked laser is to comparable our microcomb result from 1 MHz to 1 Hz. For an integration band from 1 MHz to 1 kHz, the lower floor of the mode-locked laser system results in only 0.6 fs of jitter, compared to 1 fs for the microcomb.

Further simplification and improvement of the overall system is attainable by implementing recent achievements on compact cavity-stabilized CW lasers. Self-injection locking[27], modulation-free locking[28,29], and opto-electronic oscillator laser locks[30,31] can support high-fidelity locking of compact diode lasers to ultrastable cavities. Moreover, on-chip frequency references and miniature in-vacuum bonded cavities can reduce both the size and housing requirements of the cavity reference[32,33]. The combination of feedforward correction of the noise from a compact comb source with such CW laser sources represents a compelling path towards fully integrated, photonics-based low noise microwave systems. This work is a significant step toward this goal.

**Methods**

*Frequency comb setup details*

For the solid-state laser experiment, the output of the 10 GHz mode-locked laser was amplified from 30 mW to 140 mW with polarization maintaining (PM) erbium-doped fiber amplifier (EDFA), and split into three branches with fiber couplers for microwave generation and detection of the two heterodyne beat notes. The frequencies of these heterodyne beat notes with the 1550 nm and 1560 nm CW lasers are around 6.7 GHz and 3.8 GHz, respectively. The beat signals are amplified and mixed to produce a signal at 2.88 GHz. A commercial digital frequency divider then divides the mixed heterodyne signal output of ~2.88 GHz by a factor of 58. To overcome the additive phase noise floor of -153 dBc/Hz of the commercial divider[34], a purpose-built regenerative divider (divide by 2, for a total division ratio of 116) was implemented after the commercial divider. This reduces the achievable phase noise floor to ~-159 dBc/Hz. The frequency of the feedforward correction signal emerging from the last divider stage is ~24.7 MHz. The remaining branch from the EDFA illuminated the photodiode for the 10 GHz microwave generation. The average optical power on the 10 GHz photodiode was ~6 mW, with a corresponding photocurrent of ~3 mA. This photocurrent was chosen to minimize the photodetector's amplitude-modulation (AM) to phase-modulation (PM) conversion[35]. The 10 GHz signal is amplified to >7 dBm by a low-phase noise amplifier (additive phase noise <-170dBc/Hz at 10kHz) to drive the local oscillator (LO) port of the second mixer. The amplified microwave is then mixed with the divider output, resulting in a microwave carrier at ~9.977 GHz. This low noise, feedforward-corrected microwave signal is then directed to a phase noise measurement system for analysis.

For micro-comb experiment, the biggest difference is the output power of 100 µW that was also amplified to 55 mW by the same EDFA. The frequencies of heterodyne beats with the 1550 nm and 1560 nm CW lasers were around 2.2 GHz and 2.5 GHz, respectively. The CW lasers are in exactly the same condition (same cavity modes, same parameters for PDH stabilization). To obtain the correct $N \times \delta f_{rep}$ signal, the sum of the beat frequencies around 4.70 GHz was taken from the mixer. Since the microcomb repetition rate is roughly twice that of the solid-state laser, the total division factor was 58. For the microcomb, a regenerative divider was not necessary, since the phase noise floor was limited by optical noise on the 20 GHz microwave (due to low power seeding the EDFA) and not the commercial divider. The photocurrent for the 20 GHz microwave generation was also ~3 mA with the same photodetector. Although it was not at an AM-to-PM node for the 20 GHz carrier, AM-to-PM conversion was not a significant contributor to the phase noise.

*Residual noise of CW laser stabilization*

For 2p-OFD with a 1 THz bandwidth frequency comb, the relative noise between the two lasers is divided by 40 dB when generating a 10 GHz carrier. When the two lasers are locked to a common cavity, the relative noise consists of the residual noise of the individual lasers, the noise added by the locking servos, and the residual cavity noise. For a 10 GHz carrier, common-mode rejection of the cavity length fluctuations reduces the residual cavity noise another ~40 dB, such that total rejection of the cavity noise approaches 80 dB [36]. This leaves the residual laser noise and the locking servo noise as the dominant contributions. Therefore, high gain, broadband feedback to the laser as well as a high signal-to-noise-ratio in the feedback servo loop are very important to attain low noise with any 2p-OFD architecture, including our feedforward method. This section discusses the phase noise contributions from our cavity-stabilized lasers.

Two CW lasers centered at 1550 nm and 1560 nm are stabilized to the common compact ULE cavity as shown in Fig. 2. The cavity FSR is about 24 GHz with the linewidth of ~30 kHz at 1550 nm. The two CW lasers are combined through a polarization maintaining dense wavelength division multiplexer, coupled to the cavity, and reflected to a common photodetector for PDH error signal generation. The input power at 1550 nm and 1560 nm to the cavity was about 400 µW and 800 µW, respectively. As the two CW lasers are modulated by EOMs with the different frequencies of ~24 and ~82 MHz, we can extract and separate each PDH error signal for feedback stabilization.

The in-loop residual phase noise of both PDH stabilizations is shown in Extended Data Fig. 1. Also shown is the phase noise limit imposed by the finite SNR of the PDH error signals[37,38]. The total contribution of the cavity-stabilized reference lasers to the microwave output is the quadrature sum of these four traces (we assume the noise of the separate locks is uncorrelated). The residual in-loop phase noise is limited by the free-running laser noise and the gain of the PDH servo loop. The SNR is limited by shot noise at the PDH detector and electrical noise of the post-detector amplifiers, resulting in a white frequency floor (20 dB/decade slope in phase noise). The PDH feedback sometimes overcompensates the error signal, such that residual in-loop goes lower than the SNR limit. However, the true out-of-loop noise floor of the PDH lock cannot go below this SNR limitation. Thanks to the combination of AOM and EOM feedback on each laser, both servo bumps (seen on the in-loop residual traces) are well-suppressed below -120 dBc/Hz on the optical carrier. To the best of our knowledge, this is record-low in-loop residual phase noise of PDH stabilization, and shows the great potential of 1 THz-span 2p-OFD to achieve below -160 dBc/Hz at >1 kHz and -180 dBc/Hz at 10 kHz on a 10 GHz microwave. Further improvements will require higher optical power and/or an optical reference cavity with a higher quality factor.

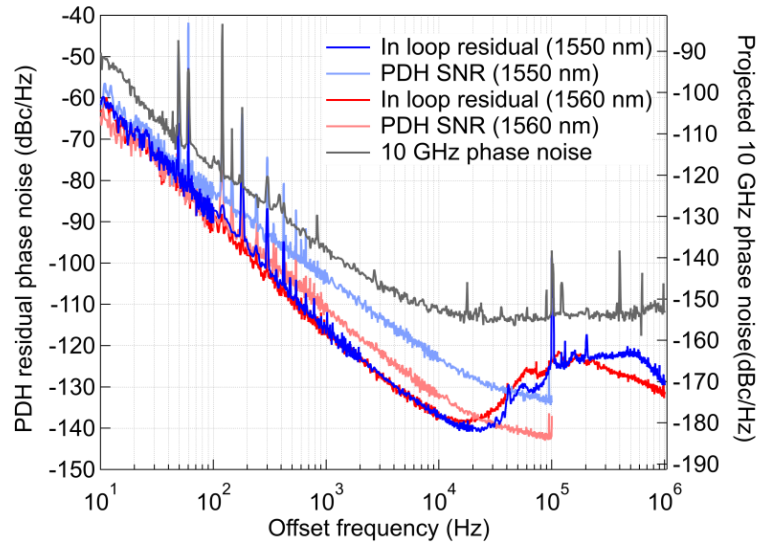

Extended Data Fig. 1 **In-loop residual phase noise of PDH locking.** Measured in-loop residual phase noise of the 1550 nm laser (blue) and the 1560 nm laser (red). Measured PDH SNR (shot noise limited) at 1550 nm (thin red) and 1560 nm (thin blue). The right axis shows the projected 10 GHz phase noise with 1.16 THz-span 2p-OFD. The 10 GHz phase noise with the solid-state laser is also shown for comparison (gray).

*Capability and limitation of the feedforward method*

Beyond solid-state mode-locked lasers and microcombs, in principle the proposed method should work with any frequency comb source, such as an electro-optic comb or semiconductor diode mode-locked laser, as long as the comb source has a broad enough spectrum and high enough optical power. For comb sources with high repetition rate noise, the electrical divider must faithfully divide the noise and the relative delay between the directly photodetected microwave signal and the divider output must be managed. Here we discuss limitations on the divider and the delay.

To verify the capability of the divider to handle high-noise signals, a 5 GHz tone from an RF synthesizer that is strongly frequency-modulated by white noise was used as a proxy for $N \times \delta f_{rep}$ with a much nosier comb source. The microwave spectra before and after frequency division (divided by 58) are shown in Extended Data Fig. 2(a). The steep noise drop-off on the divided signal is due to the limited ~10 MHz bandwidth of the noise modulation. With the noisy signal input, the divider still reduces the phase noise as expected, as determined by the following calculation. White frequency noise modulation of the 5 GHz tone results in a measured 20 MHz linewidth. For white frequency modulation, there is a simple relationship between the linewidth and the noise power spectral density, given by $\Delta \nu = \pi S_\nu$, where $\Delta \nu$ is the full width at half maximum linewidth and $S_\nu$ is the frequency noise in units $Hz^2/Hz$[39]. This frequency noise level is converted to phase noise, and then compared to the measured phase noise after frequency division. Extended Data Fig. 2 (b) shows excellent agreement between predicted and measured phase noise values, verifying that the divider faithfully divides such a noisy signal. Also shown in Extended Data Fig. 2(b) is the free-running microcomb phase noise from the main text, ~40 dB lower at 10 kHz offset than the noise-modulated synthesizer. Therefore, the divider should support most comb sources available today.

Even with high-fidelity noise division, a relative delay between the photodetected microwave and the feedforward correction signal will limit the amount of the noise suppression. Since the delay of the divider is very low, matching the temporal delay should be manageable with a compact system. The level of phase noise suppression at 10 kHz offset as a function of delay is shown in Extended Data Fig. 2(c). A relative timing error of less than 1 ns (about 20 cm of RF cable or optical fiber) should be sufficient for even the highest noise comb sources, supporting 83 dB noise suppression at 10 kHz. The calculated phase noise with the noisy 5 GHz tone combined with feedforward with 1 ns delay is shown in Extended Data Fig. 2(d), indicating this level on a comb source has the possibility of reaching <-130 dBc/Hz on a 10 GHz carrier. To achieve the similar noise suppression with a conventional feedback scheme, a net feedback loop response time, including the actuator of the comb and the loop filter, needs to be at the 1 ns level. Moreover, to achieve 80 dB noise suppression at 10 kHz, an integrator feedback circuit and comb actuator would require > 100 MHz bandwidth. This is well beyond the capabilities of frequency comb feedback servos. For further suppression, careful attention to non-uniform group delay of any bandpass filters in the feedforward path is necessary. The measured response of the tunable bandpass filter we used in solid-state mode-locked laser can be approximated as a 4th order 33 MHz Butterworth filter around 2.8 GHz. This should support >100 dB suppression at 10 kHz in our calculation. Demonstrating large noise suppression with other comb sources will be the subject of future investigations.

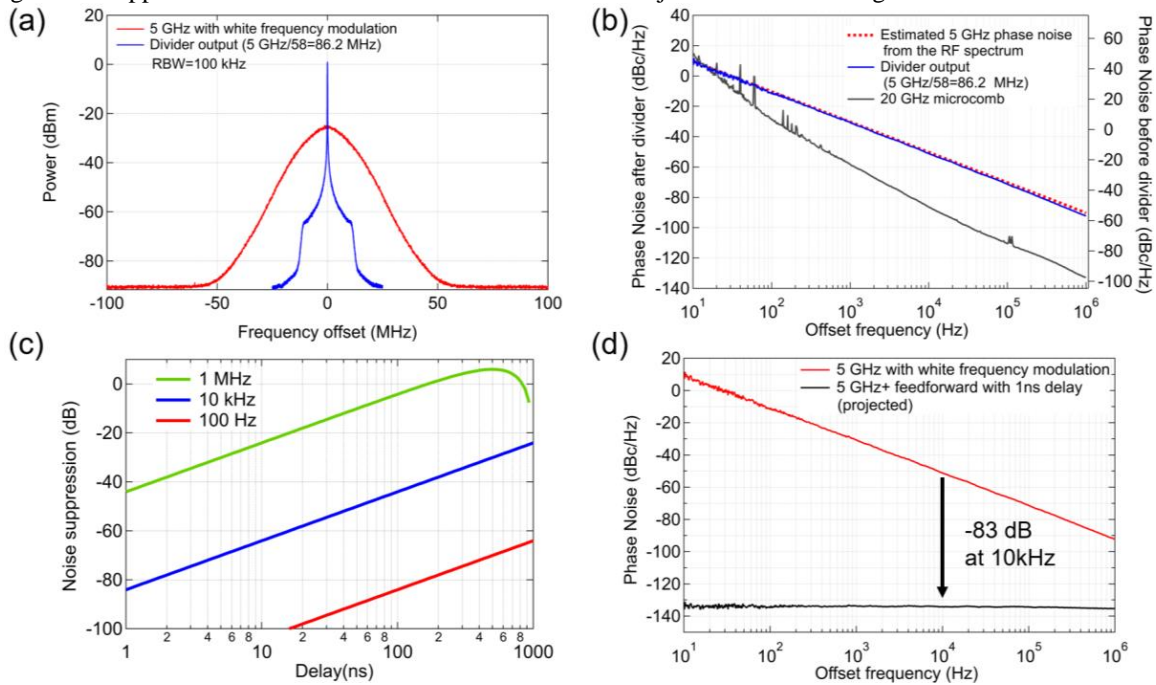

Extended Data Fig. 2 **Assessing divider capability towards nosier comb sources.** (a) RF spectra of noisy 5 GHz source (red) and the corresponding divider output at 86.2 MHz (blue). The 5 GHz signal imitates the mixer output of $N \times \delta f_{rep}$ with a much nosier comb. (b) Phase noise comparison between noisy 5 GHz tone estimated from 3dB bandwidth (dotted red), measured phase noise after the divider (blue), and measured phase noise of the 20 GHz microcomb (gray). (c) Calculated noise suppression with different delays. (d) Estimated phase noise with a nosier comb and 1 ns delay.

*Acknowledgment*

We thank Andrew Ludlow and the NIST Yb optical clock team for ultrastable reference light, Fabrizio R. Giorgetta, Charles A. McLemore, and Dahyeon Lee for helpful comments on the paper. Product names are given for scientific purposes only and do not represent an endorsement by NIST. The research reported here performed by A. Savchenkov, V. Iltchenko, W. Zhang, and A. Matsko was carried out at the Jet Propulsion Laboratory at the California Institute of Technology, under a contract with the National Aeronautics and Space Administration (80NM0018D0004). Funding provided DARPA and NIST.